# Wavelets, Curvelets and Multiresolution Analysis Techniques Applied to Implosion Symmetry Characterization of ICF Targets

Bedros Afeyan,[a,**] Kirk Won,[a] Jean Luc Starck,[b] and Michael Cuneo[c]

[a] Polymath Research Inc., Pleasanton, CA
[b] Centre d'Etude de Saclay, CEA, France
[c] Sandia National Laboratories, Albuquerque, NM
** e-mail: bedros@polymath-usa.com

## ABSTRACT

*We introduce wavelets, curvelets and multiresolution analysis techniques to assess the symmetry of X ray driven imploding shells in ICF targets. After denoising X ray backlighting produced images, we determine the Shell Thickness Averaged Radius (STAR) of maximum density, $r^*(N, \theta)$, where N is the percentage of the shell thickness over which to average. The non-uniformities of $r^*(N, \theta)$ are quantified by a Legendre polynomial decomposition in angle, $\theta$. Undecimated wavelet decompositions outperform decimated ones in denoising and both are surpassed by the curvelet transform. In each case, hard thresholding based on noise modeling is used. We have also applied combined wavelet and curvelet filter techniques with variational minimization as a way to select the significant coefficients. Gains are minimal over curvelets alone in the images we have analyzed.*

## 1. INTRODUCTION

High yield targets for inertial confinement fusion (ICF) have been designed recently which rely on a double Z pinch configuration.[1] These novel approaches to indirect drive inertial confinement fusion energy production promise to attain fusion by next generation Z pinch machines such as those proposed by Sandia National Laboratories.[2] The idea is to attain the uniform implosion of an annular shell containing fusion fuel via intense X ray illumination in an intermediate chamber sitting between two wire array Z pinches which create the necessary mega-Joules of X rays.[1-3] To diagnose the imploding shell at the center of the middle chamber of a double Z pinch hohlraum (DZPH), an X ray backlighter is used.[4] The X rays must enter the middle chamber from its side, and image it through a slit on the other side of the chamber at an external film plane.[4] The implosion is due to the radiation driven ablation of the shell driven by copious amounts of X rays that enter the central chamber having been created by the two imploding Z pinch wire arrays on either side, which generate mega-Joules of X rays each.[1-3] These X ray fluxes, if sufficiently synchronized to allow for a uniform radiation distribution inside the central hohlraum chamber, will drive the uniform implosion of the fusion capsule in its center. The quantitative assessment of the degree to which this is so is our goal when we denoise such X ray backlighting generated images. The three images we will denoise and analyze are given in Fig. 1.

Ideally, one would acquire a sequence of such images per implosion using gated optics and track the evolution of a given accelerating shell.[4] Distortions in the shell as it implodes can be due to non-uniformities in the X ray illumination, target surface imperfections and their amplification due to hydrodynamic instabilities such as Rayleigh-Taylor.[1] From a given X ray backlighting image, one would like to extract the degree of asymmetry of the imploding shell near its peak plasma density or, equivalently, its radius of minimum X ray transmission.[5] Identifying that radius is the task at hand with appropriate shell thickness averaging and denoising. The main sources of the noise are thought to be the graininess of the film and the scanner digitization process.[3-5] This suggests that the noise is additive and the empirical evidence is that it is white Gaussian. For the very symmetric shot, Z927, and the intermediate one, Z926, this is certainly the case. For Z928, however, the noisiest and most distorted image analyzed herein (since it is taken later in time and therefore further along in the compression and thus has a much smaller radius) the log of the image was used and a Gaussian noise model seemed to fit that better. The techniques utilized in this paper should find fruitful application in the denoising of X ray backlit laser driven ICF target implosion images from facilities such as Omega[6] and the NIF.[7]

## 2. WAVELETS, CURVELETS AND DENOISING

In the past fifteen years, there has been a revolution in signal processing by the introduction and popularization of multiresolution analysis techniques and in particular, those based on wavelets.[8-9] The reign of Fourier as the dominant spectral representation domain where filtering, smoothing, scale identification and even phase space tiling concepts are evoked has been challenged by the advent of wavelets and their fast algorithmic implementation.[8-10] The successes of wavelets and their usage are too many to review here. A visit to www.wavelets.org will reveal the history of, tutorials on, and latest news concerning, wavelets and their uses from astronomy to medicine, pure mathematics to audio engineering. For the purposes of this paper, the important tools are wavelets for image processing,[10] curvelets for image processing[11-12] and noise modeling (statistical analysis) and discrimination/thresholding techniques.[10-13]

The fundamental concept of multiresolution analysis is to decompose a signal or image into time (or space) and scale *simultaneously*. Thus, just as in music notation, not only which notes are to be played are specified (what Fourier supplies very precisely) but *when* and for what *durations* (about which Fourier is silent). The simultaneous scale and time (or space) decompositions are carried out in the simplest case by equispaced translates (whose number per scale or spacing is scale dependent) and $2^J$ dilates of a mother wavelet[8-9] (where J is the number of scales to be used). A wavelet is a localized waveform with zero mean and good localization properties both in space (time) and reciprocal space (frequency) (i.e. good localization in phase space). In signal processing terms, a wavelet is a band pass filter around a low pass filter used to detect the coarsest features of a signal (which itself is known as the scaling function in wavelet parlance). Wavelets record the finer and finer scale structures around the coarsest scale ones in a nested set of levels, with proportionately more translates at each finer scale. This pyramid structure can be constructed by the basic fast algorithm (O(N) operations, N being the number of data points, FFT being O(N lnN)) that was invented in the late eighties.[8-9] The resulting orthogonal or biorthogonal decompositions are said to be *decimated wavelet* ones. The drawback in denoising applications is that these decompositions are not translationally invariant while noise presumably is (how would the noise know which bit to corrupt?). The scale information content of a signal is the same no matter when we start or end the signal just as long as we keep all of it (all sequential permutations of the bits, that is). To adhere to that symmetry, the number of translates at each multiresolution level (scale) should not be changed and at all scales the number of translates should be the same as that at the finest scale. This gives rise to a highly redundant representation which nevertheless has great advantages in denoising since now the noise is diluted in a great many more coefficients only the most significant of which is to be kept via hard thresholding.[10-14] Denoising is done by estimating the type and variance of the noise and choosing to keep large wavelet coefficients which have a low probability of being noise. This is done iteratively as described in numerous references.[10-15]

The three images in Fig. 1 will be denoised using undecimated and decimated wavelets as well as curvelets. We have used the variance of the noise found in the individual images to estimate the coefficients which are likely to be noise dominated and discarded them. This is referred to hard thresholding.[10-15] The decimated wavelet transform, being a non-redundant transform, does not allow for optimal denoising since it is not translationally invariant in its construction. Fig. 2 shows denoising using a $5\sigma$ iterative denoising method using the Antonini 9/7 biorthogonal set of wavelets.[10] Decimated biorthogonal wavelet transforms such as these are far more useful for image compression than for denoising.[8-9]

Fig. 3 shows undecimated wavelet transforms with hard thresholding based on noise modeling.[10,13,15] This highly redundant transform does allow for optimal denoising in the wavelet domain since it is translationally invariant in its construction. Discrimination against noise is much easier in this transform than in the decimated transform case. However, many artifacts still remain since a point wise and isotropic construction is being implemented when wavelets are used. To adapt to the contours of the figure a truly 2D construction is needed such as that afforded by curvelets.[11-12] Curvelets afford a very useful tool for denoising images such as these. Fig. 4 shows this directly.

One way to see how the imploding shells behave as a function of angle is to plot the locus of the radius of *minimum* X ray transmission sandwiched between the loci of the inner and outer radii of *maximum* transmission. These max-min-max polar plots are shown in Fig. 5 for the curvelet transformed and denoised images. The asymmetry of the implosion can be easily seen from the nonuniformity of the distance separating the various curves. Z927 is the most symmetric while Z928 is the least.

A better way to characterize the deviations of a shell from spherical symmetry (in its projection onto a plane) is to unwrap the circular ribbon like structures by plotting the images as radius vs. angle on a Cartesian grid. Now a circular shell becomes a ribbon, a circle becomes a horizontal line and any deviations from that line imply distortions and fluctuations. Nestled in the middle of the ribbons, we also plot $r^*(0, \theta)$, $r^*(50, \theta)$ and $r^*(90, \theta)$ where the first argument of $r^*$ is the percentage of the distance between the minimum transmission radius and its closest maximum (on either side of the minimum). This is the shell thickness averaged radius (STAR), $r^*$, where $I(r, \theta)$ is the X ray intensity distribution in the image:

$$r^*(N,\theta) = \int_{r_L(N,\theta)}^{r_R(N,\theta)} I(r,\theta) r \, dr \bigg/ \int_{r_L(N,\theta)}^{r_R(N,\theta)} I(r,\theta) \, dr$$

$$I(r_L) = I(r_R) = \frac{N}{100} \left( Min[I(r_{max,L}), I(r_{max,R})] - I(r_{min}) \right).$$

We have found that $r^*(90, \theta)$ has the smoothest and most useful qualities of the three shown in Figs 6.

Next, in Fig. 7, we plot the $r^*(90, \theta)$ curves (each normalized with its own average radius), obtained by curvelet filtering. For Z928, only curvelets produce a smooth $r^*(90, \theta)$ curve. Finally, the Legendre polynomial decompositions of curvelet generated $r^*(90, \theta)$ curves for Z926-Z928 are given in Figure 8. The degrees of asymmetry in the imploding shells are thus quantified. The vertical axes contain the Legendre polynomial coefficients $c_n$ normalized to the zero order Legendre polynomial coefficient $c_0$. $c_0$ is equal to the average diameter of the curve $r^*(90, \theta)$. The coefficients $c_n$ are found by Legendre decomposing each half of $r^*$ separately and adding the results. For further details see Ref. 14-15. Combined filtering promoted in Ref 12, for instance, was seen in our case not to improve the results obtained with curvelets alone.[15]

## ACKNOWLEDGMENTS

This work was supported by a contract with Sandia National Laboratories. The authors would like to thank Ken Struve, Dillon McDaniel, John Porter, Jim Hammer, Roger Vesey, Guy Bennett, Dan Sinars and Nino Landen for their valuable input and encouragement.

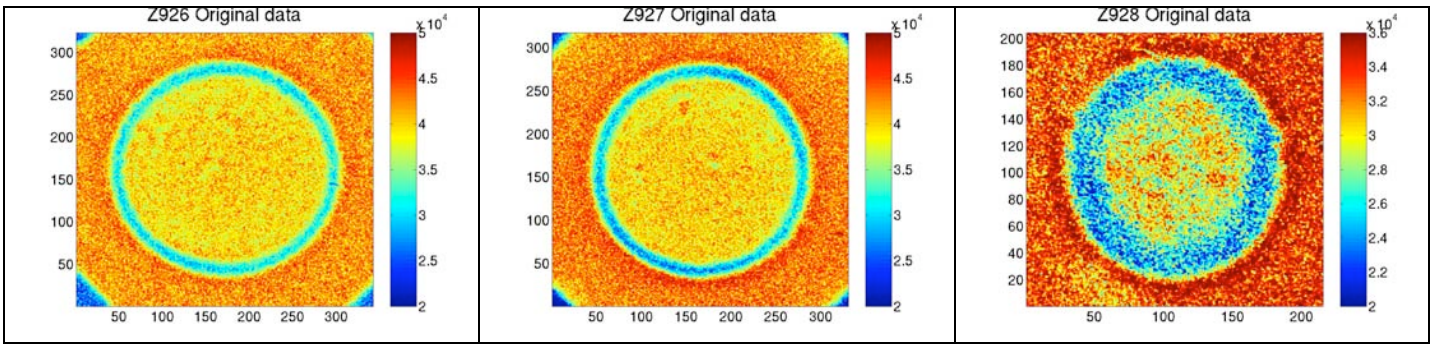

**Figure 1.** X Ray backlighting generated images of double Z pinch compressed hollow shells in Sandia Z machine shots Z926-Z928. The axes are in pixels. The conversion factor is 5 microns per pixel. Originally, before compression, the shells had millimeter size radii.

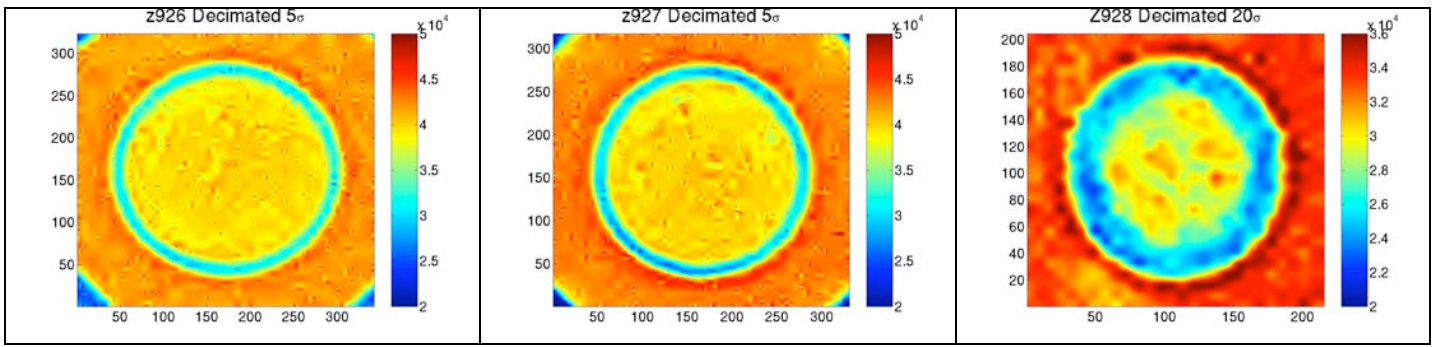

**Figure 2.** Decimated Wavelet transform reconstructions of X Ray backlighting images.

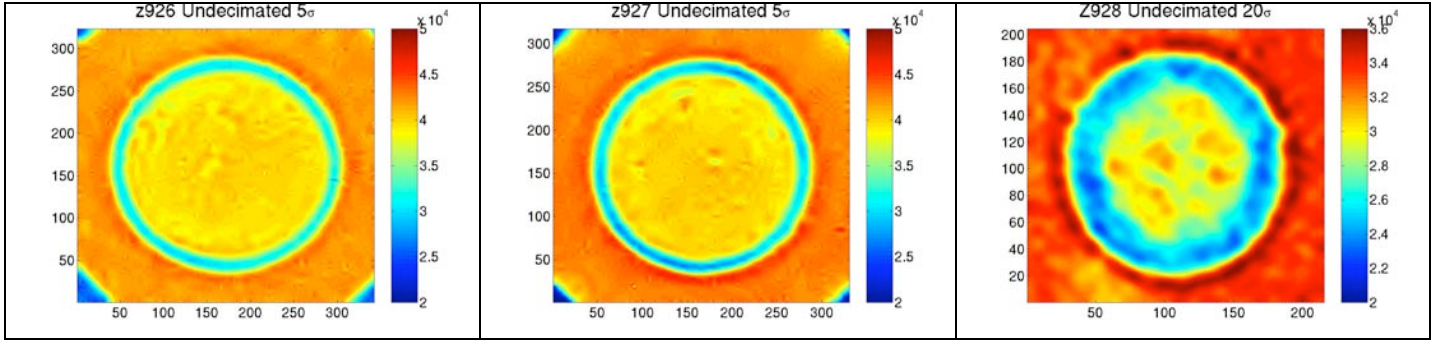

**Figure 3.** Undecimated Wavelet transform reconstructions of X Ray backlighting images.

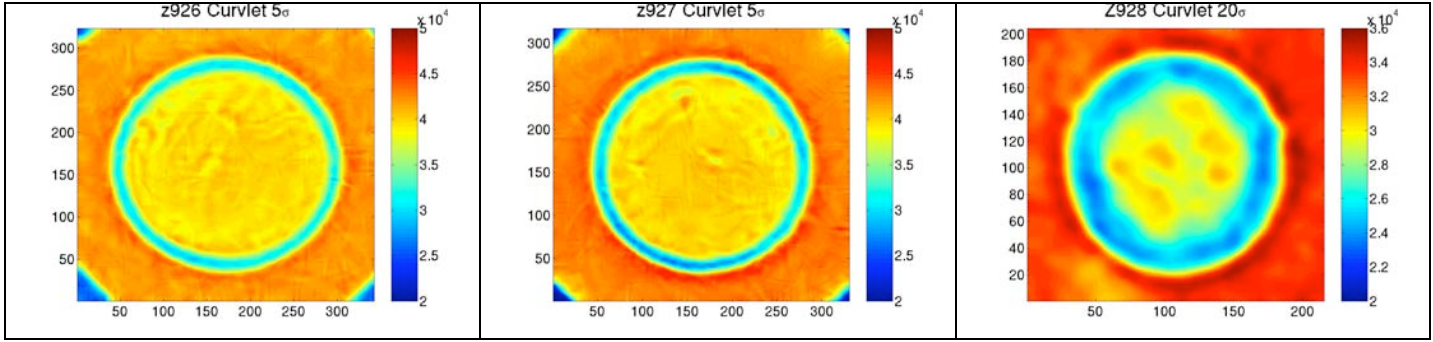

**Figure 4.** Curvelet transform reconstructions of X Ray backlighting images.

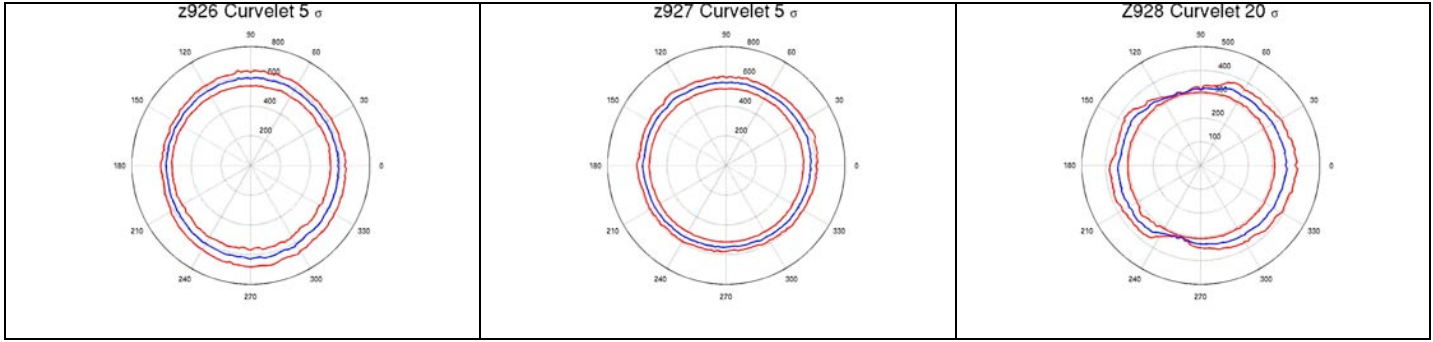

**Figure 5.** Polar max-min-max plots of minimum and maximum X ray transmission radii for curvelet transform reconstructed X ray backlighting images of Z926-Z928. The reason the Z928 min and max curves seem to pinch around a 100 degrees and again around 250 is because the shells seem to be flattened there blurring the concept of min and max.

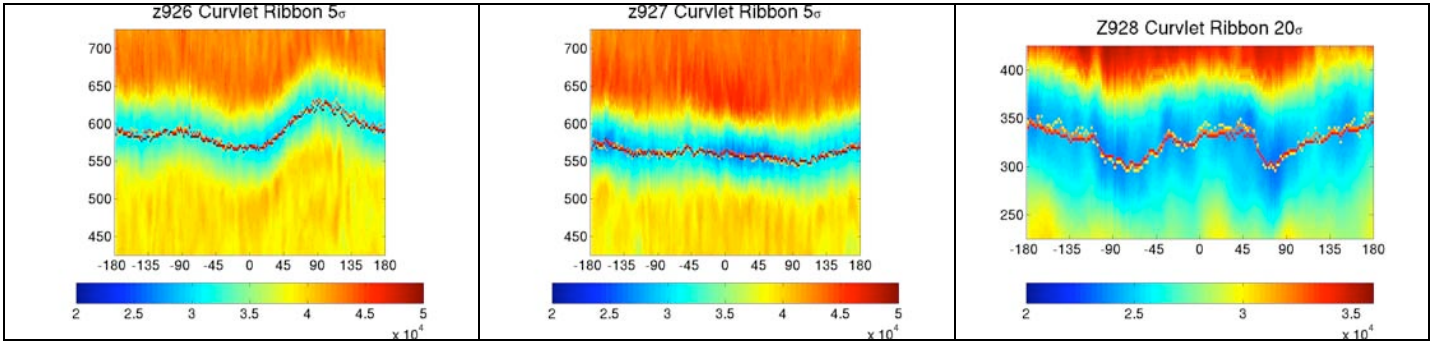

**Figure 6.** Interpolated polar coordinates ribbon plots: Curvelet transform reconstructions of X Ray backlighting image of double Z Pinch target shots Z926 with $r^*(0, \theta)$, $r^*(50, \theta)$ and $r^*(90, \theta)$ superposed.

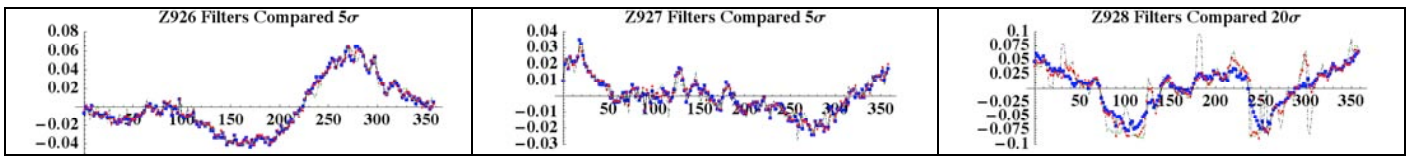

**Figure 7.** A comparison of the $r^*(90, \theta)$ curves obtained via the three filters on the three data sets each normalized to its own average radius. The average radii for the three Z images are roughly 603, 574 and 340, respectively. Black (solid lines) corresponds to a decimated wavelet decomposition, red (diamonds), undecimated, and blue (squares) corresponds to a curvelets based decomposition. Note that only curvelets can detect the $R^*(90)$ curves for Z928. Wavelets fly off the mark.

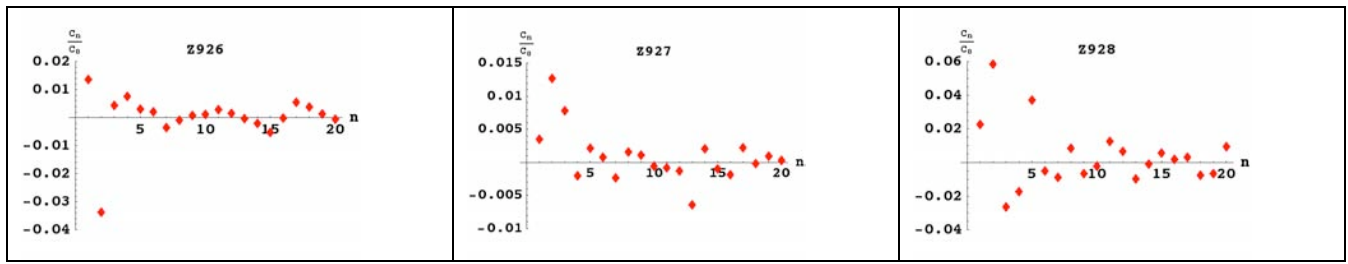

**Figure 8.** The Legendre polynomial decomposition of curvelet generated $r^*(90)$ curves for Z926-Z928. The first 20 mode amplitudes are shown. We see that Z927 has around 1.5% $P_2$ and 0.2% $P_4$ deformations, Z926 has over 3.5% $P_2$ and almost 1% $P_4$ deformations while Z928 has over 6% $P_2$ and 2% $P_4$ deformations. These are decompositions without any attempts at centering the images to within 5 pixels of the true centers. That is why $P_1$ is nonzero but small in all three cases.